\documentstyle[preprint,aps,epsf]{revtex}

\begin{document}

\title{Pipe network model for scaling of dynamic interfaces in porous media}

\author{Chi-Hang Lam$^{1}$ and Viktor K. Horv\'{a}th$^2$} 

\address{$^1$Department of Applied Physics, Hong Kong Polytechnic
University, Hung Hom, Hong Kong\\ 
$^2$Department of Physics, University of Pittsburgh, Pittsburgh, 
Pennsylvania 15260,\\
and 
Dept. of Biological Physics,
E\"{o}tv\"{o}s Univ., P\'azm\'any P. 1A. 1117 Budapest, Hungary}

\date{\today}
\maketitle

\begin{abstract}
We present a numerical study on the dynamics of imbibition fronts in
porous media using a pipe network model.  This model quantitatively
reproduces the anomalous scaling behavior found in imbibition
experiments [Phys. Rev. E {\bf 52}, 5166 (1995)].  Using simple
scaling arguments, we derive a new identity among the scaling
exponents in agreement with the experimental results.
\end{abstract}

\pacs{PACS numbers: 05.40.+j, 47.55.Mh, 05.70.Ln, 68.35.Fx}

Self-affine interfaces are found in a variety of phenomena
such as two-phase flow in porous media, thin film deposition, flame
fronts, etc. \cite{Barabasi}. In particular, scaling properties of
imbibition fronts have been the focus in many experimental studies
\cite{Buldyrev,Horvath,Amaral}. Static scaling
behavior has been suggested to be described by a directed depinning
percolation model \cite{Buldyrev,Tang,Amaral2,Albert}. To explain
dynamic properties, one need to go
beyond local models and properly deal with the effects of long-range
coupling through pressure. Much progress has been obtained using
theoretical approaches including consideration of capillary waves
\cite{Flekkoy}, a Flory-type scaling argument \cite{Ganesan} 
and very recently a phase-field model \cite{Dube}.
These studies contributed to our better understanding of the scaling
behavior of imbibition.  However, exponents provided in general do
$not$ compare satisfactorily with the experimental values. In fact, of the
four exponents determined experimentally by Horv\'{a}th and Stanley
\cite{Horvath} with good precision, $none$ has previously been
explained analytically or numerically, and no existing exponent
identity applies. In this work, we successfully reproduce all these
exponents using a pipe network model once the relevant model
parameters are properly fine tuned.  A new exponent identity among
these exponents is also presented.  In addition, we have found an
empirical relation between the spatial and temporal correlations.

Horv\'{a}th and Stanley studied imbibition fronts climbing up in
vertical filter paper sheets which move continuously downward into a
water container at constant speed $v$ \cite{Horvath}.  The interface
height at horizontal coordinate $x$ and time $t$ is denoted by
$h(x,t)$.  The average height $\overline{h}$ and 
width $w$ were demonstrated to scale as
\begin{eqnarray}
  \label{Omega}
  \label{thetaL}
  v \sim \overline{h}^{-\Omega}   \mbox{~~~~and~~~~}   w \sim v^{-\kappa} 
\end{eqnarray}
The temporal height-height correlation function
$C(t)$=\/$<[h(x,t'+t) - h(x,t')]^2>^{1/2}$
\cite{temp_corr} was shown to follow a scaling form,
\begin{equation}
  \label{cvt}
  C(t) = v^{-\kappa} f (t\/v^{(\theta_{t}+\kappa)/\beta}),
\end{equation}
where $f(u)\sim u^{\beta}$ for small $u$, and it 
converges to a constant
for large $u$. The exponents were found to be
\begin{equation}
   \label{exponents}
   \Omega=1.594 \mbox{~,~}  \kappa=0.48 \mbox{~,~} 
   \theta_{t}=0.37  \mbox{~,~~~and~} \beta=0.56 .
\end{equation}
The first exponent has particularly strong implications
because Darcy's law, which is well established for many porous media
\cite{Sahimi}, implies $\Omega=1$.
The same result was also found in numerous
numerical studies \cite{Dube,Aker,Blunt}.  
Physically, the anomalous
value $\Omega=1.594$ suggests that as the interface height increases, its
propagation speed becomes slower than simple considerations of
capillary pressure and viscous drag would suggest. 
This cannot be explained by gravity, evaporation,
deformation of wetted paper \cite{Dube}, or inertial effects
\cite{Sahimi} since they involve intrinsic length or time scales and
can only introduce transients or cut-offs to Darcy's prediction. 

Our network model consists of cylindrical pipes connecting volumeless
nodes on a two-dimensional square lattice with periodic boundary
conditions in the horizontal direction. According to Possiuelle's law
\cite{Sahimi}, the flow rate in a pipe of radius $r$ is $Q={\pi
r^4}\Delta p/{8\mu d}$ where $\Delta p$ is the pressure difference
between two points separated by $d$.  The fluid viscosity is denoted
by $\mu$. The pressure at the water level in the tank is maintained at
the atmospheric value. The capillary pressure is ${2\gamma} \cos
\phi/r$ in each partially filled pipe where $\gamma$ and $\phi$ are
the surface tension and the contact angle respectively. Gravity is
neglected. This is because in Ref. \cite{Horvath}, scalings were
reported only for $\overline{h}$ smaller than 65\% of the maximum
value limited by gravity at $v=0$ \cite{Horvath2}. We have checked
numerically that gravity is negligible in this range, even though it
can lead abruptly to the breakdown of Eq. (\ref{Omega}) at larger
$\overline{h}$ and finally to the pinning transition.
We calculate the pressure at every wetted node above the water level
of the tank by solving Kirchoff's equations using a locally adaptive
over-relaxation method. The propagation of the interface during
the associated adaptive time step is hence computed. For
computational convenience, possible deviation from Possiuelle's law at
the junction of the pipes is neglected as usual \cite{Aker} and
menisci cannot retreat after reaching a node.  Since in real situation
air can escape from either side of the sheet, trapping of air is
irrelevant and its flow is not simulated.

The pipe model at this point is similar to those in previous studies
\cite{Sahimi,Aker,Blunt} most of which focused on flow inside porous rocks.
Tenuous percolation type wetting patterns are obtained. To simulate
more compact patterns observed in paper, we therefore implement a new
wettability rule to impose a strong effective surface tension in our
network. Now, water can enter an empty pipe from a wetted node only if
the neighborhood is sufficiently wet. Specifically, of the 8 most
nearby nodes (the 4 nearest and the 4 next nearest neighbors) forming
a loop around the node under consideration, there must exist a
connected subset of 5 which are already wet. This is the most
stringent criterion for the wettability of pipes connected to a wetted
node without halting the imbibition entirely or involving
consideration of further neighbors. It leads to a very strong
effective surface tension. Different and more detailed wettability
rules were discussed in Ref. \cite{Blunt}. In addition, fluid pathways
in randomly arranged fibers have complicated geometry and there are
abundance of narrow bottlenecks and large pores. We therefore propose
another important characteristic of our network, namely large spatial
fluctuations in the local properties. Its implementation will be
discussed later.

The model was first tested in the presence of small fluctuations.
Adopt a unit lattice spacing and specify the fluid properties by
$\mu=\gamma\cos \phi =1$ \cite{transform}.  We assume random pipe
radii in the range [0,0.5] uniformly distributed
\cite{radius}.  This distribution is broader than those used by others
\cite{Aker,Blunt}. The Darcy's prediction $\Omega=1$ is readily verified.

\begin{figure}
   \epsfbox[80 100 800 600]{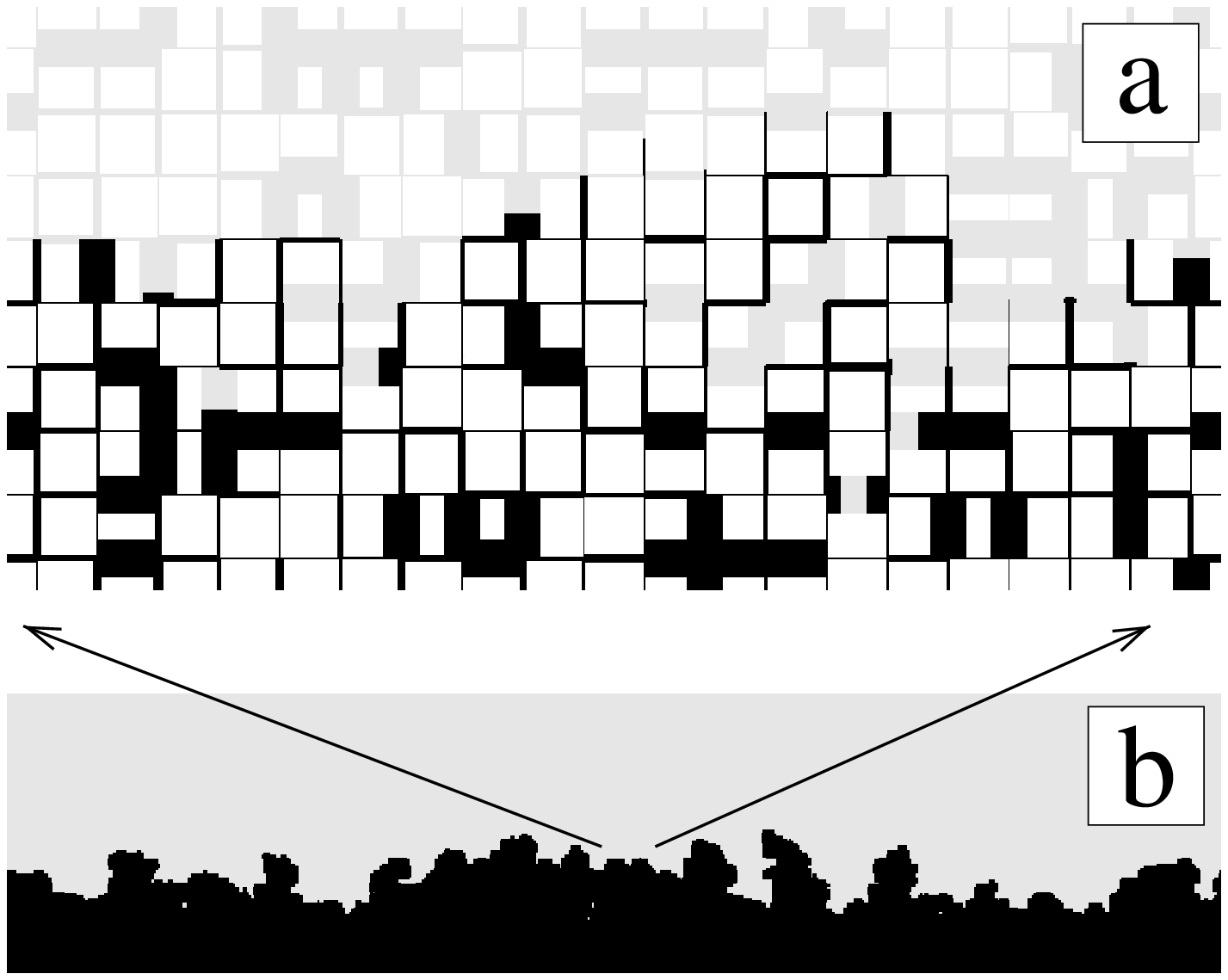}
   \caption{\narrowtext (a) Snapshot of a small scale simulation
   showing individual pipes.  Wet and dry regions are shaded in black
   and grey respectively.  Radii of pipes have been scaled down. (b)
   Snapshot of the wetted region in a larger scale simulation on a
   lattice of width 500.  The average height is $36.8$. The arrows
   provided only to demonstrate the ratio of the scales of the two
   figures.}  \label{Fnetwork}
\end{figure}

To simulate large fluctuations expected in real paper, we simply
replace a fraction $p_w$ of pipes by wider ones with radius $r_w$. For
sufficiently large $p_w$ and $r_w$, $\Omega>1$ is obtained. Both
exponents $\Omega$ and $\kappa$ then depend non-trivially but
continuously on $p_w$ and $r_w$. Fortunately, we have found that nice
power-laws with the experimental values of $\Omega$ and $\kappa$ can
be reproduced if we put $p_w=0.18$ and $r_w=2.0$ \cite{radius}. We
will adopt these parameters in the remainder of this
work\cite{somewhereelse}.  Snapshots of simulations of imbibition in
stationary sheets are shown in Fig. \ref{Fnetwork}.

\begin{figure}
   \epsfbox[80 100 800 600]{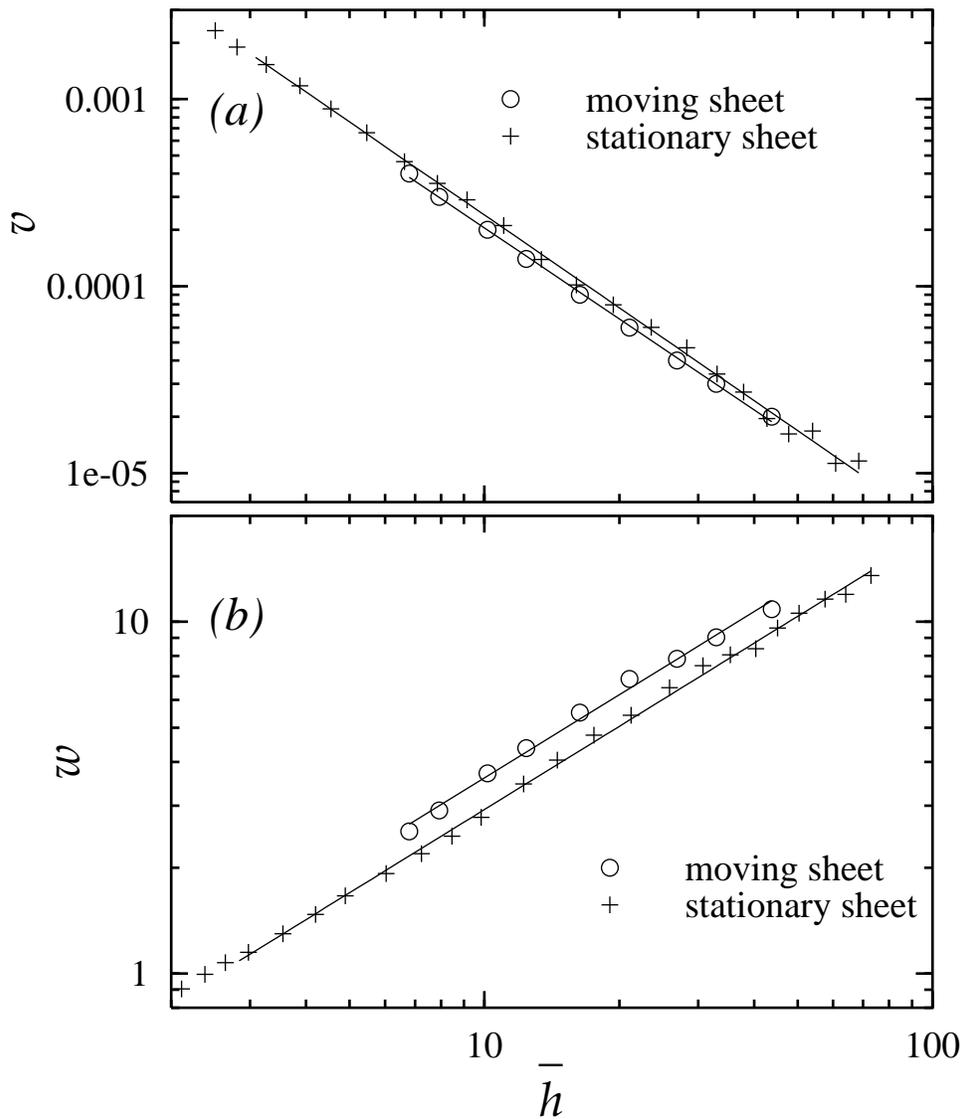}
   \caption{\narrowtext Log-log plots of (c) $v$ and (b) $w$ against
   $\overline{h}$.  In (a), the fitted lines have slopes -1.62 and
   -1.65 for the moving and stationary case respectively. In (b), they
   are 0.785 and 0.786.}  \label{FVH}
\end{figure}

We have simulated imbibition in moving paper sheets considered in
Ref. \cite{Horvath}. This involves continuous upward shifting of the
level on the lattice which represents the contact line of the paper
sheet with the water in the reservoir. The interface height
${\overline{h}}$ and width $w$ are measured after steady state has
been attained and are averaged over 3 independent runs using lattices
of width 200 and height 1000.  Figures \ref{FVH}(a) and (b) plot
respectively $v$ and $w$ against $\overline{h}$. The linearity
observed in log-log plots verifies the scaling relations in
Eq. (\ref{Omega}). We get $\Omega=1.62 \pm 0.05$ and $\kappa=0.49 \pm
0.03$ close to the experimental values in Eq. (\ref{exponents}) due to
the fine tuning of the pipe distribution mentioned above. The quoted
errors represent only the uncertainties in the linear fits.

\begin{figure}
   \epsfysize 8in   \epsfbox{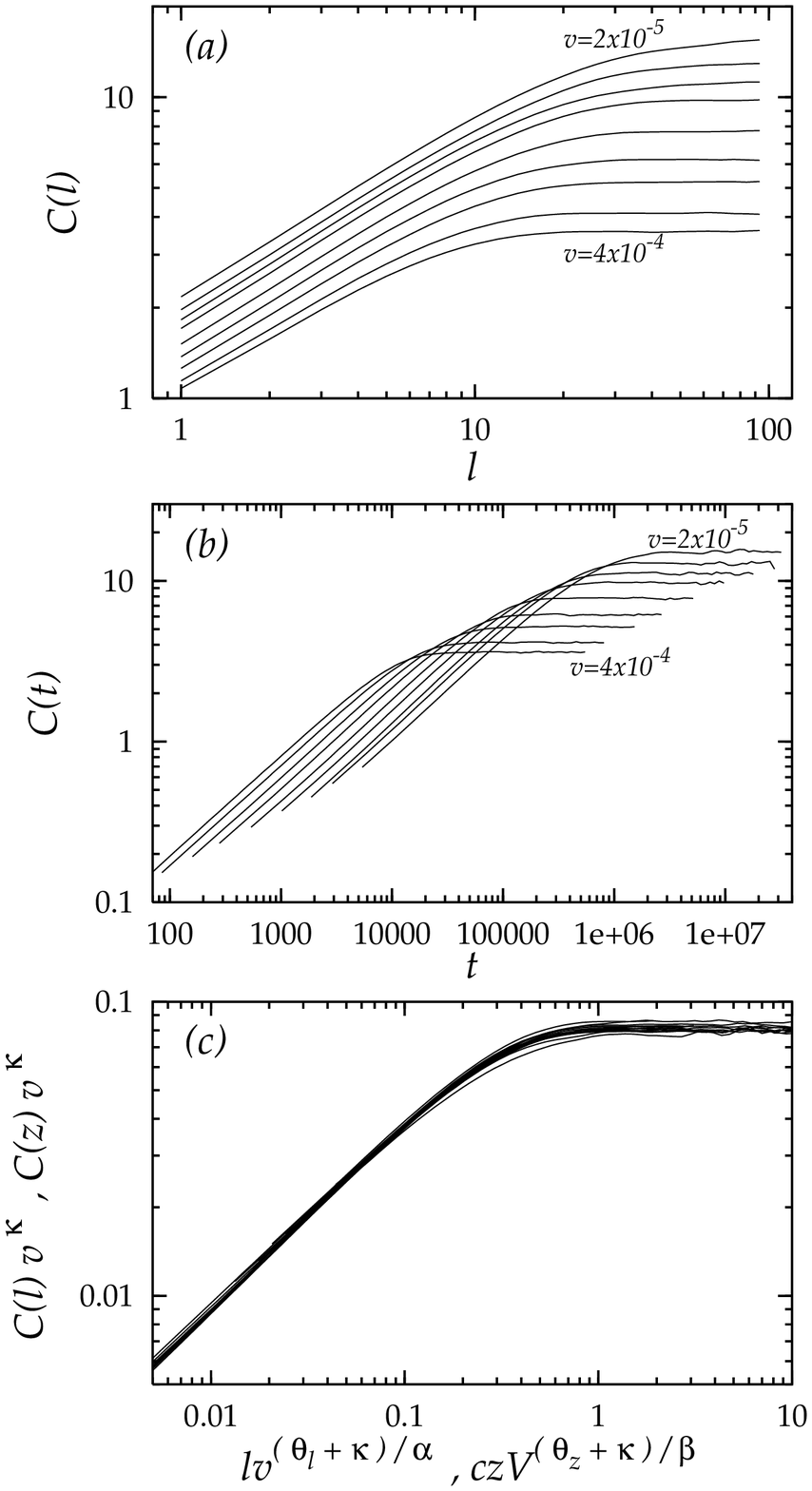}
   \caption{\narrowtext Log-log plots of (a) spatial correlation, (b)
   temporal correlation and (c) simultaneous data collapse of all
   correlation curves in both (a) and (b) for paper speed $v$ given by
   $v/10^{-5}=2,3,4,6,9,14,20,30$ and 40.}  \label{FCr}
\end{figure}

The spatial correlation $C(l)$=$<$$[h(x$+$l,t)$-$h(x,t)]^2$$>^{1/2}$
and the temporal counterpart $C(t)$ defined earlier are computed and
plotted in Figs. \ref{FCr}(a) and (b) respectively.  The initial
linear regions corresponding to $C(l)\sim l^\alpha$ and $C(t)\sim
t^{\beta}$ for small $l$ and $t$ give $\alpha=0.61 \pm 0.01$ and
$\beta=0.63 \pm 0.01$ respectively. This value of $\beta$ is in reasonable
agreement with the experimental value 0.56.
We have verified that $C(t)$ in Fig. \ref{FCr}(b) follows the
experimentally motivated scaling forms in Eq. (\ref{cvt}) and we
obtain $\theta_t=0.38 \pm 0.02$ in excellent agreement with the
experimental value 0.37.  Similarly, the spatial correlation follows
an analogous scaling form
$
  C(l) = v^{-\kappa} g(l\/\/v^{(\theta_l+\kappa)/\alpha})
$
where $g(u)\sim u^\alpha$ for small $u$ and it
converges to a constant for large $u$. 
A further result is obtained
by first reparametrizing $C(t)$ by the vertical displacement $z=v\/t$
of the sheet. Equation (\ref{cvt}) then becomes
$
  C(z) = v^{-\kappa} f(z\/v^{(\theta_z+\kappa)/\beta})
$,
where $\theta_z=\theta_t - \beta$. 
We note that both $C(l)$ and $C(z)$ can be collapsed individually
using a common set of exponents $\alpha=\beta=0.62$ and
$\theta_l=\theta_z=-0.24$.
Furthermore, they can all be collapsed together as shown in
Fig. \ref{FCr}(c) verifying the empirical identities
\begin{eqnarray}
  \label{id3}
  \alpha=\beta \mbox{~~,~~~} \theta_l=\theta_z \mbox{~~~~and~~~~}
  f(cu)=g(u)
\end{eqnarray}
where $c=1.6$.

Next, we simulate imbibition in stationary paper sheets. The 
interface height $\overline{h}$ and width $w$ averaged over 7
lattices of width $1000$ are found to scale as
\begin{equation}
  \label{epsilon}
   {\overline{h}} \sim t^\epsilon \mbox{~~},
   \mbox{~~~and~~}
   w \sim t^{\beta_w}
\end{equation}
where $\epsilon=0.374 \pm 0.005$ and $\beta_w=0.29 \pm 0.01$.  The
interface speed $v(t)={d\overline{h}}/{dt}$ hence computed from direct
numerical differentiation on our data is plotted as a function of
$\overline{h}$ in Fig. \ref{FVH}(a). The width $w$ is also plotted
against $\overline{h}$ in Fig. \ref{FVH}(b). Direct comparison with
the moving sheet results is possible because the paper speed $v$ in
that case is also the average interface speed in the paper
frame. These plots show that the scaling relations in
Eq. (\ref{Omega}) apply to stationary sheets with the same exponents
as well. Therefore, we can derive Eq. (\ref{Omega}) from
Eq. (\ref{epsilon}) and vice versa. In the derivation, one obtains
\begin{equation}
   \label{id1}
   \epsilon = \frac{1}{\Omega+1}
   \mbox{~~~~and~~~~}
   \beta_w=\frac{\Omega\kappa}{\Omega+1} ,
\end{equation}
which describe our exponents accurately. 

For a stationary sheet, $w$ at a given $\overline{h}$ is only about
80\% of the corresponding value for a moving sheet as observed from
Figs. \ref{FVH}(b). The roughness has therefore not yet fully
developed for the given height $\overline{h}$. In contrast, for the
less noisy case the roughness has saturated completely and can be
determined solely from $\overline{h}$ \cite{Dube}. More importantly,
our result indicates that the roughness gets neither closer to nor
further from saturation as $\overline{h}$ increases since $w$ is
always 80\% of the saturated value.  This constancy of the degree of
saturation implies that the system has a unique dynamic time scale
dictating both the growth of the roughness and the steady state
dynamics of a roughened interface.  The time it takes to develop a
width $w$ is $w^{1/\beta_w}$ from Eq. (\ref{epsilon}). It is thus
proportional to the relaxation time extracted from the temporal
correlation function $v^{-(\theta_t+\kappa)/\beta}
\sim w^{(\theta_t+\kappa)/(\beta\kappa)} $
from Eqs. (\ref{cvt}) and (\ref{thetaL}). Therefore,
$1/\beta_w=(\theta_t+\kappa)/(\beta\kappa)$. Applying Eq. (\ref{id1}),
we have
\begin{equation}
   \label{id2}
   \beta = \Omega(\theta_t+\kappa)/ (\Omega+1) .
\end{equation}
This identity is particularly important because all values have been
measured experimentally in Ref. \cite{Horvath}. Inserting the
experimental values into the r.h.s. of Eq. (\ref{id2}), we get
$\beta=0.52$ in reasonable agreement with the experimental value
$0.56$. Using our numerical estimates instead, we get $\beta=0.54$
which is a little smaller than the numerically found value 0.61. In
the later case we relate the discrepancy to errors in the numerical
determination of $\beta$.  This may be due to lattice discretization
effects which becomes more important at $C(t) \alt 1$. The spatial
counterpart $\alpha$ should hence suffer a similar
problem. Furthermore, noticeably different values of $\alpha$ and
$\beta$ are obtained if correlations of higher moments are used
\cite{Dube}.

The good agreement between our numerical results and the experimental
ones strongly supports that our model has captured the essential
physics in the imbibition process. However, some further points are
yet to be considered.  First, despite the nice power-law fits and data
collapses of the relevant quantities observed, existence of
extraordinarily slow crossover effects masking different asymptotic
scalings should be cautioned. Second, our model reproduces the
experimental results after fine tuning of the radius distribution. We
have also found a few other completely different distributions which
also lead to similar exponents upon fine
tuning\cite{somewhereelse}. The exponents nevertheless are $not$
robust with respect to changes in the details of the models. Using
distributions other than the fine tuned ones, the exponents are in
general different and sometimes scalings may not even
hold\cite{somewhereelse}. We expect that there is a yet unknown
selection mechanism so that models generating the experimental
exponents are preferred. This is currently under active
investigations. However, one cannot determine $apriori$ a correct
realistic pipe distribution. This is because all pipe networks are
simplified models of fluid pathways which are indeed very different
from those of paper. Realistic simulations, for example, using
lattice-Boltzmann method with sophisticated boundary conditions in 3
dimensions
\cite{Koponen} unfortunately covers only microscopic regions and thus
cannot be applied to study the macroscopic scalings.

Finally, local models of imbibition have been classified into
isotropic and anisotropic universality classes exemplified
respectively by the random field Ising model (RFIM) and the directed
percolation depinning (DPD) model \cite{Amaral2}. The anisotropy in
the DPD model is due to a solid-on-solid condition. In contrast, for
our model both the wettability rule and the flow dynamics treat the
horizontal and vertical directions equivalently. Concerning the local
symmetry, it is more closely related to the isotropic class. It was
found that $\alpha=1$ and $\beta_w=3/4$ for the isotropic class
\cite{Amaral2} while $\epsilon=\alpha=\beta_w\simeq 0.633$ for the
anisotropic one
\cite{Tang} at criticality. This is to be compared with
$\epsilon\simeq 0.374$, $\alpha\simeq 0.61$ and $\beta_w \simeq 0.29$
for our model. It is easy to see that the much larger values of
$\epsilon$ and $\beta_w$ for DPD and also $\beta_w$ for RFIM are due to
the locality of the interections. We believe that the agreement of
$\alpha$ between DPD and our model is just a coincidence. The
morphologies of the surfaces are indeed visually quite
different. Previous works on non-local models including pipe networks
and the phase-field model
\cite{Dube,Aker,Blunt} are in a weak fluctuations regime and are all
consistent only with Darcy's law. Our pipe network with strong
fluctuations is the only model exhibiting distinctly different
scalings in good agreement with the experimental ones in
Ref. \cite{Horvath}. We therefore suggest that it belongs to a new
universality class.

In conclusion, we have simulated imbibition in paper using a pipe
network model and reproduced all scaling behaviors observed
experimentally in Ref. \cite{Horvath}. We obtain 
\begin{equation}
   \Omega=1.62 \mbox{~,~}  \kappa=0.49  
   \mbox{~,~} \theta_{t}=0.38 \mbox{~~~and~~} \beta=0.63 
\end{equation}
which is to be compared with Eq. (\ref{exponents}).  The model
displays rich behaviors, and can be tuned to reproduce the
experimentally determined values of $\Omega$ and $\kappa$. Then
$\beta$ and $\theta_t$ turn out respectively in reasonable and
excellent agreement with the experimental values. Assuming a single
time scale in the dynamics, we have presented a new exponent identity
(Eq. (\ref{id2})) which is justified by the experimental values. Two
other exponent identities in Eq. (\ref{id1}) relating the moving and
stationary paper cases are deduced and verified numerically. Further
identities in Eq. (\ref{id3}) for exponents and scaling functions are
suggested empirically based on a data collapse between the spatial and
temporal correlations.

We have benefitted from helpful communications with L.M. Sander, M
Rost, M. Dub\`e, T. Ala-Nissila and F.G. Shin who are gratefully
acknowledged. C.H.L. is supported by project no. B-Q075 from Research
Grants Council of Hong Kong SAR.  V.K.H is supported by the Hungarian
Science Foundation grant OTKA-F17310 and NATO grant DGE-9804461.

\begin{references}


\bibitem{Barabasi}
A.-L. Barab\'{a}si and H.E. Stanley, Fractal concepts in Surface
Growth, Cambridge University Press (1995).

\bibitem{Buldyrev}
S. V. Buldyrev et al.,
Phys. Rev. A {\bf 45}, R8313 (1992);

\bibitem{Horvath}
V. K. Horv\'ath and H. E. Stanley, Phys. Rev. E {\bf 52},
5166 (1995)

\bibitem{Amaral}
L. A. N. Amaral et al.,
Phys. Rev. Lett. {\bf 72}, 641 (1994);
F. Family, K. C. B. Chan, and J. G. Amar, in Surface Disordering,
Roughening and Phase Transitions (Nova Science, Commack, NY, 1992)
p.205; T. H. Kwon, A. E. Hopkins, and S. E. O'donnell, Phys. Rev. E
{\bf 54} 685 (1996);
O. Zik et al., 
Euro. Phys. Lett. {\bf 38} 509 (1997).

\bibitem{Tang}
L.-H. Tang and H. Leschhorn, Phys. Rev. A {\bf 45}, R8309 (1992).

\bibitem{Amaral2} L. A. N. Amaral, A.-L. Barab\'asi, and H. E. Stanley,
Phys. Rev. Lett. {\bf 73}, 62 (1994);
L-H. Tang, M. Kardar and D. Dhar,
Phys. Rev. Lett. {\bf 74}, 920 (1995). 

\bibitem{Albert}
R. Albert, A.-L. Barab\'{a}si, N. Carle, and A. Dougherty,
Phys. Rev. Lett. {\bf 81}, 2926 (1998).

\bibitem{Flekkoy}
E. G. Flekk{\o}y and D. H. Rothman, Phys. Rev. Lett. {\bf 75}, 260 (1995).

\bibitem{Ganesan}
V. Ganesan and H. Brenner, Phys. Rev. Lett. {\bf 81}, 578 (1998).

\bibitem{Dube}
M. Dub\`e et al.,
Phys. Rev. Lett. {\bf 83}, 1628 (1999).

\bibitem{temp_corr} 
More precisely,
$C(t)=\left<[\tilde{h}(x,t')-\tilde{h}(x,t'+t)]^2\right>^{1/2}$ in
Ref. \cite{Horvath}, where $\tilde{h}(x,t)=h(x,t)-{\overline
h(t)}$. The two definitions are equivalent for wide paper sheets or
pipe networks since the spatially averaged height $\overline
h(t)\equiv \overline h$ is then time independent at steady state. For
narrower networks used in our simulations, our definition suffers
smaller finite size effects and can better approximate the results in
Ref. \cite{Horvath} where wide sheets were used.

\bibitem{Sahimi}
M. Sahimi, Flow and transport in porous media and fractured rock, VCH
(1995).


\bibitem{Aker}
E. Aker and K. J. M{\aa}l{\o}y, and A. Hansen, Phys. Rev. E {\bf 58}, 2217 (1998).

\bibitem{Blunt}
M. J. Blunt and H. Scher, Phys. Rev. E {\bf 52}, 6387 (1995).

\bibitem{Horvath2} V. K. Horv\'ath, unpublished.


\bibitem{transform}
From dimensional analysis, this amount to
transforming length, time and fluid density with the respective
multiplicative factors $a_0^{-1}$, $ a_0^{-1} \mu_0^{-1} \gamma_0
\cos\phi$ and $ a_ 0 \mu_0^{-2} \gamma_0\cos \phi$ respectively. Here,
$a_0$, $\mu_0$ and $\gamma_0$ are the lattice spacing, viscosity and
surface tension respectively in physical units.

\bibitem{radius}
According to the dynamics we have defined, enlarging or reducing the
radii of $all$ pipes by an overall multiplicative factor is only
equivalent to a transformation of units similar to that in
Ref. \cite{transform} and is of no physical significance. Possible
physical overlapping of pipes is hence irrelevant and is naturally
neglected in our dynamical equations.

\bibitem{somewhereelse}
Results on other pipe distributions will be presented elsewhere.

\bibitem{Koponen}
A. Koponen et al.,
Phys. Rev. Lett. {\bf 80}, 716 (1998).
\end{references}
\end{document}